\documentclass[twocolumn,epjc3]{svjour3}  

\pdfoutput=1

\smartqed 

\RequirePackage{fix-cm}
\RequirePackage{graphicx}
\usepackage{epsfig,float}
\usepackage{amsfonts}
\usepackage{amsmath}
\usepackage{longtable}
\usepackage[colorlinks=true,linkcolor=red,citecolor=blue]{hyperref}
\usepackage{wrapfig}
\usepackage{enumitem}
\usepackage{nicefrac}
\usepackage[parfill]{parskip}
\usepackage{orcidlink}
\usepackage[frozencache,cachedir=minted-cache]{minted}
\usepackage{booktabs}
\usepackage{multirow}
\usepackage{array}
\usepackage{cite}
\newcommand{\PreserveBackslash}[1]{\let\temp=\\#1\let\\=\temp}
\newcolumntype{C}[1]{>{\PreserveBackslash\centering}p{#1}}
\newcolumntype{R}[1]{>{\PreserveBackslash\raggedleft}p{#1}}
\newcolumntype{L}[1]{>{\PreserveBackslash\raggedright}p{#1}}

\setminted{breaklines=true}

\journalname{Eur. Phys. J. C}

\begin{document}

\title{Open database for GPD analyses}

\author{
V.D.~Burkert\orcidlink{0000-0003-4747-0838}\thanksref{addrJLAB}
\and
A.~Camsonne\orcidlink{0000-0003-4333-2614}\thanksref{addrJLAB}
\and
P.~Chatagnon\orcidlink{0000-0002-4705-9582}\thanksref{addrJLAB,addrCEA}
\and
K.~Cichy\orcidlink{0000-0002-5705-3256}\thanksref{addrUAM}
\and
M.~Constantinou\orcidlink{0000-0002-6988-1745}\thanksref{addrTemple}
\and
H.~Dutrieux\orcidlink{0000-0001-8334-4885}\thanksref{addrWaM}
\and
I.~M.~Higuera-Angulo\orcidlink{0000-0002-5600-8875}\thanksref{addrJLAB}
\and
C.~Mezrag\orcidlink{0000-0001-8678-4085}\thanksref{addrCEA}
\and
D.~Richards\orcidlink{0000-0001-6971-873X}\thanksref{addrJLAB}
P.~Sznajder\orcidlink{0000-0002-2684-803X}\thanksref{addrNCBJ}
}

\institute{
Thomas Jefferson National Accelerator Facility, 12000 Jefferson Avenue, Newport News, VA 23606, USA \label{addrJLAB}
\and
~Irfu, CEA, Universit\'e Paris-Saclay, F-91191 Gif-sur-Yvette, France \label{addrCEA}
\and
~Faculty of Physics and Astronomy, Adam Mickiewicz University, ul.\ Uniwersytetu Pozna\'nskiego 2, 61-614 Pozna\'n, Poland \label{addrUAM}
\and
~Department of Physics, Temple University, Philadelphia, PA 19122 - 1801, USA \label{addrTemple}
\and
~Physics Department, William \& Mary, Williamsburg, VA 23187, USA \label{addrWaM}
\and
~National Centre for Nuclear Research, NCBJ, 02-093 Warsaw, Poland \label{addrNCBJ}
}

\date{\today}

\sloppy

\maketitle

\begin{abstract}
This article summarizes the main ideas behind creating an open database proposed for use in the exploration of generalized parton distributions (GPDs). This lightweight database is well suited for GPD phenomenology and is designed to store both experimental and lattice-QCD data. It can also aid in benchmarking GPD-related developments, such as GPD models. The database utilizes a new data format based on the YAML serialization language, enabling the storage of essential information for modern analyses, such as replica values. It includes interfaces for both Python and C++, allowing straightforward integration with analysis codes.
\end{abstract}

\section{Introduction}
\label{intro}

Generalized parton distributions (GPDs)~\cite{Muller:1994ses,Ji:1996nm,Radyushkin:1996nd} arise in the factorisation theorems of perturbative quantum chromodynamics. These universal objects describe the non-perturbative parts of exclusive reactions, such as deeply virtual Compton scattering (DVCS)~\cite{Ji:1996nm,Belitsky:2001ns}, and are associated with the partonic structure of hadrons. To some extent, GPDs can be considered a unification of parton distribution functions (PDFs) and elastic form factors, which are typically studied without any connection. The importance of GPDs in particle physics arises from their relation to nucleon tomography~\cite{Burkardt:2002ks,Burkardt:2002hr}, allowing the study of the position of partons carrying a specific fraction of a hadron’s momentum, as well as their connection to the energy-momentum tensor~\cite{Ji:1996ek,Polyakov:2018zvc}. This latter feature enables the determination of the total angular momentum carried by specific partons and provides insight into the so-called ``mechanical properties'' of media composed of confined partons, see e.g. \cite{Burkert:2018bqq,Burkert:2023wzr,Duran:2022xag,Burkert:2023atx}. The phenomenology of GPDs, compared to that of PDFs, comes with additional difficulties. The most striking challenge arises from the three-dimensional nature of GPDs, meaning that more data are required to constrain these objects than in the case of one-dimensional PDFs. The situation is worsened by the variety of GPD types that simultaneously contribute to exclusive reactions, defined for combinations of hadron-parton spins -- some of which do not even have counterparts in inclusive physics. Exclusive reactions also have relatively small cross-sections, and the requirement of reconstructing all particle states imposes non-trivial constraints on the experimental apparatus.

Despite these difficulties, a collection of data has already been measured by experiments at JLAB, DESY, and CERN (for a review, see for instance Ref.~\cite{Kumericki:2016ehc,dHose:2016mda}). These data are not only for DVCS but also for processes like deeply virtual meson production (DVMP)~\cite{Collins:1996fb,Passek-K:2023nbe}, timelike Compton scattering (TCS)~\cite{Berger:2001xd,CLAS:2021lky}, and exclusive heavy meson production~\cite{Ivanov:2004vd,Flett:2021ghh}. Several other exclusive processes have been proposed in recent years, including double deeply virtual Compton scattering (DDVCS)~\cite{Belitsky:2002tf,Guidal:2002kt,Deja:2023ahc} and $2 \to 3$ processes~\cite{Duplancic:2018bum,Boussarie:2016qop,Grocholski:2021man,Grocholski:2022rqj,Qiu:2024mny}. All of the aforementioned processes, measured under various experimental conditions (such as beam and target polarization states and different particles), provide complementary information about GPDs and help to pinpoint their specific contributions. It is also important to mention that the GPD topic will be a pillar of the physics program of current and future QCD laboratories, particularly JLab~\cite{Accardi:2023chb}, EIC~\cite{AbdulKhalek:2021gbh}, and EIcC~\cite{Anderle:2021wcy}.

An additional source of GPD information is provided by lattice QCD computations, traditionally via moments of these objects extracted from local matrix elements (see, e.g., Refs.~\cite{LHPC:2007blg,Bali:2018zgl,Alexandrou:2020sml,Shintani:2018ozy,Alexandrou:2022dtc,Alexandrou:2021wzv,Alexandrou:2023qbg}) or, more recently, through the novel techniques utilizing non-local matrix elements, the quasi- ~\cite{Ji:2013dva} and pseudo-distribution methods \cite{Radyushkin:2017cyf} or the off-forward Compton amplitude approach \cite{Hannaford-Gunn:2024aix}.
These recent techniques have already witnessed an extensive amount of work aimed at several types of the nucleon's GPDs, see e.g.~\cite{Alexandrou:2020zbe,Alexandrou:2021bbo,Bhattacharya:2022aob,Cichy:2023dgk,Bhattacharya:2023ays,Bhattacharya:2023nmv,Holligan:2023jqh,Bhattacharya:2024qpp,HadStruc:2024rix,Hannaford-Gunn:2024aix,Bhattacharya:2024wtg}. 
It is expected that lattice QCD results will be crucial in constraining GPDs in the phase space that are inaccessible (or practically difficult to access) through currently measured exclusive processes or those to be measured in the foreseeable future. The limited access to GPDs through some exclusive processes is discussed in the literature under the topic of ``shadow GPDs''~\cite{Bertone:2021yyz,Dutrieux:2021wll,Moffat:2023svr,Riberdy:2023awf,Cichy:2024afd}. 

Despite the wealth of existing data sensitive to GPDs -- a collection that we expect to grow significantly in the future due to upcoming experiments and lattice QCD calculations -- the community still lacks an open database that could accelerate research progress. Such a centralized platform providing easy access to GPD-related data could serve several purposes. The most straightforward benefit would be saving the time needed to incorporate existing and new data into phenomenological analyses. Several groups employing various modeling strategies are currently working on global analyses of GPDs (see, for instance, Refs.~\cite{Kumericki:2015lhb,Moutarde:2018kwr,Guo:2023ahv, Kriesten:2020apm}). Each of these groups could benefit from using the same database, directly fed by experimental and lattice QCD collaborations. Another advantage of using a common database would be improved reproducibility of phenomenological analyses, enabling the community to fully adopt open-science standards. Additionally, the proposed database could also be populated with pseudo-data, not only those used in impact studies but also those serving as benchmarks for theoretical developments, such as values obtained from new GPD models. This would allow others to verify that their implementation of these models is correct.

We stress that the proposed database has been designed based on the experience of GPD phenomenologists and lattice-QCD practitioners and, therefore, fully addresses the needs of future analyses aimed at exploring parton distributions. It also provides direct integration with existing codes and, in general, is much better tailored for GPD physics than other open databases used in particle physics, in particular HEPData~\cite{hepdata}, which is designed for the general purpose of storing data in high-energy physics.

In this article, we summarise the main ideas for the open database we propose for use in the GPD community. We begin with Sect.~\ref{sec:basics}, which describes the essential whys and hows behind the project, including the server selection, format, programming languages, etc. In Sect.~\ref{sec:data_format}, we detail the data format we have adopted for this purpose. This format allows for storing both experimental and lattice QCD data and includes many useful features, such as the ability to store bin boundaries and replica values. Finally, in Sect.~\ref{sec:ui}, we discuss the database interface that users can integrate into their codes, without providing too many technical details, which can be accessed elsewhere, specifically on the project's website~\cite{database}. A concise summary is provided in Sect.~\ref{sec:summary}.


\section{Basics}
\label{sec:basics}

The project aims to create and maintain a database capable of storing a collection of experimental and lattice-QCD data. A specific format is used to store such data, and a dedicated library is provided in both C++ and Python, allowing users to easily access the database in their analysis codes. Additionally, a webpage~\cite{database} has been created, featuring up-to-date documentation and a list of available data.

To achieve this ambitious goal despite limited computing and human-power resources, we utilized GitHub~\cite{github}, one of the basic tools used in the particle physics community. This choice offers several advantages. By using an existing service, we eliminate the need to maintain a separate database server or host a webpage. GitHub provides free hosting, with the option to mirror the project, e.g., on local GitLab instances. This helps prevent situations where access to the database is lost, such as when a university discontinues its hosting service or an administrator leaves the field. GitHub also allows the formation of a management team and offers a straightforward interface for interacting with users. These features can be used to report issues, suggest improvements, and add new data. Additionally, version control is simple and transparent. Altogether, these features enhance the chances that the database will remain useful to the community, preventing it from fading into obscurity.

The database stores files in a text format (see Sect.~\ref{sec:data_format} for details). This approach simplifies database maintenance, as each data series is stored in a separate file. Adding new data is equally simple; for example, an experimental collaboration needs only to provide a new text file in the specified format to incorporate it into the database. However, the use of text data files has a drawback arising from their size. Two factors must be considered in this context: the limitations of GitHub's service and the potential burden on users’ computing systems. The initial size of all files in the database (which includes most of the world DVCS data) is only a few megabytes. We estimate that the adopted solution should be feasible for at least several years. We do not plan to store ``intermediate'' data produced in lattice-QCD computations (which can reach gigabytes in size), opting instead to store only ``final'' data that can be used in phenomenological analyses. If using GitHub becomes impractical, we can migrate the project to a different server, use a premium account, or adopt an alternative solution, such as transferring the data to a MySQL server -- an option we considered when developing the data format.

The database is initially populated with the published experimental data (respecting the appropriate licenses) and some lattice-QCD results. The project's webpage is created using Jekyll technology and is automatically updated whenever the project is modified. For example, adding a new data file triggers the regeneration of the webpage, including an updated list of stored data. The Python and C++ interfaces are complete (see Sect.~\ref{sec:ui} for more details). The Python package is automatically deployed to the PyPi service~\cite{pypi}, while the C++ interface is a wrapper for the Python code. This allows us to write essentially one codebase for use in both programming languages. Finally, a series of automated checks have been defined on GitHub, helping to reduce the risk of releasing a broken version of the project.


\section{Data format}
\label{sec:data_format}

In this section, we describe the structure of the data files we intend to store in the database. For the sake of simplicity and portability, we have chosen the YAML data serialization format~\cite{yaml}, which can be easily processed in both Python and C++ codes. YAML is an ASCII-based format, where the structure of data is defined by indentation. An example of a dummy data file presenting all available features users can use when storing data in the database is shown below. Thanks to the use of human-readable keys and additional comments (starting with '\#'), the presented structure of data should be intuitive even for non-expert users. An additional description is provided below the example.

\begin{minted}[
    fontsize=\footnotesize,
    % linenos,
    % frame=leftline,
    % framesep=5mm,
    % tabsize=2, 
    % breaklines
  ]{yaml}

---
uuid: 9j7gof4d # id

general_info:

   date: 2023-09-25 # date of insertion to database
   data_type: DVCS # data type (enum)
   pseudodata: true # tag to distinguish between data and pseudo-data (e.g. Monte Carlo study)
        # (optional) if not set pseudodata = false
   collaboration: "Example" # collaboration name releasing data stored (limited to 40 characters)
   reference: "arXiv:01/02" # reference, (limited to 255 characters)
   conditions: # experimental or lattice-QCD conditions
      lepton_beam_type: e- # lepton beam type (enum)
      lepton_beam_energy: 10. # lepton beam energy (by default in GeV)
      hadron_beam_type: p # hadron beam type (enum)
      hadron_beam_energy: 100. # hadron beam energy (by default in GeV)
        # if missing, fixed target assumed
   comment: "Comment" # comment (optional, limited to 255 characters)

data:

    - data_set: 

        label: "Q2_dep" # label of current data set
                        # limited to 40 characters 
   
        kinematics: # data related to kinematics

            name: [xB, Q2] # definition of kinematic phase-space, 
                           # here 2D (xB, Q2) domain (enums)
            unit: [none, GeV2] # units (enums)
            value: # mean values of xB and Q2 obtained in three kinematic bins 
                - [0.2, 1.5] # 1st bin: xB = 0.2, Q2/GeV2 = 1.5 
                - [0.3, 2.2] # ...
                - [0.4, 3.1] # 3th bin: xB = 0.4, Q2/GeV2 = 3.1 
            # or alternatively: 
            # value: [[0.2, 1], [0.3, 2], [0.4, 3]]

            unc: # uncertainties associated to the mean values (optional)
                - [0.0015, 0.016] # 1st bin: xB = 0.2 +- 0.0015, Q2/GeV2 = 1.5 +- 0.016 
                - [0.0012, 0.011] # ...
                - [0.0019, 0.013] # 3th bin: xB = 0.4 +- 0.0019, Q2/GeV2 = 3.1 +- 0.013
            bin: # bin boundaries (optional)
                - [[0.11, 0.25], [1, 2]] # 1st bin: 0.11 < xB < 0.25 and 1 < Q2/GeV2 < 2
                - [[0.25, 0.35], [2, 3]] # ...
                - [[0.35, 0.52], [3, 4]] # 1st bin: 0.35 < xB < 0.52 and 3 < Q2/GeV2 < 4
            replica: # values obtained from replicas used in the related analysis (optional)
                     # in the first bin three replicas give: 
                     # xB = {0.2005, 0.1967, 0.1995} and Q2 = {1.514, 1.502, 1.487} 
                     # the same replicas in the second bin give:
                     # xB = {0.2997, 0.3005, 0.2988} and Q2 = {2.212, 2.193, 2.207} 
                - [[0.2005, 0.1967, 0.1995], [1.514, 1.502, 1.487]] 
                - [[0.2997, 0.3005, 0.2988], [2.212, 2.193, 2.207]]
                - [[0.4002, 0.4003, 0.3987], [3.094, 3.094, 3.107]]

        observable: # data related to observables

            name: [ALU, ALL] # definition of observables (enums)
                             # here, in each kinematic bin experiment measures two observables: 
                             # ALU and ALL asymmetries 
            unit: [none, none] # units (enums)
            value: # mean values of ALU and ALL obtained in three kinematic bins
                - [0.13, 0.29] # 1st bin: ALU = 0.13, ALL = 0.29
                - [0.14, 0.24] # ...
                - [0.11, 0.26] # 3th bin: ALU = 0.11, ALL = 0.26
            stat_unc: # statistical uncertainties associated to the mean values (optional)
                - [0.03, 0.08] # 1st bin: ALU = 0.13 +- 0.03, ALL = 0.29 +- 0.08
                - [0.04, 0.07] # ...
                - [0.03, 0.02] # 3th bin: ALU = 0.11 +- 0.03, ALL = 0.26 +- 0.02
            sys_unc: # systematic uncertainties associated to the mean values (optional)
                     # here we demonstrate how one can use asymmetric uncertainties 
                     # and can specify a correlation between uncertainties
                - [0.02, 0.01] # 1st bin: ALU = 0.13 +- 0.02, ALL = 0.29 +- 0.01
                - [0.01, [0.02, 0.03]] # 2nd bin: ALU = 0.14 +- 0.01, ALL = 0.24 - 0.02 + 0.03
                - ["corr_matrix1", 0.01, 0.02] 
                    # 3th bin: ALU = 0.11 +- 0.01, ALL = 0.26 +- 0.02, 
                    # uncertainties are correlated according to 
                    # 'corr_matrix1' matrix, see 'correlation' section
            sys_unc_contrib_label: ["fit", "detector"] 
                    # labels of contributions to systematic 
                    # uncertainties (optional)
                    # limited to 40 characters 
            sys_unc_contrib: # contributions to systematic uncertainties (optional)
                - [[0.015, 0.013], [0.007, 0.007]] 
                    # contributions to systematic uncertainties of 
                    # asymmetries measured in the first bin, 
                    # i.e. 0.02 and 0.01 values (see above)
                    # contributions are:
                    # 0.015 and 0.007 for source labeled as 'fit'
                    # 0.013 and 0.007 for source labeled as 'detector'
                - [[0.007, 0.008], [[0.010, 0.017], [0.020, 0.023]]] # use of asymmetric uncertainties 
                - ["corr_matrix2", 0.005, 0.008, 0.014, 0.016] 
                    # use of correlation matrix with elements:
                    # sigma_sys_ALU_fit 
                    # sigma_sys_ALU_detector
                    # sigma_sys_ALL_fit 
                    # sigma_sys_ALL_detector
            replica: # values obtained from replicas used in the related analysis (optional)
                     # in the first bin three replicas give: 
                     # ALU = {0.132, 0.127, 0.140} and ALL = {0.295, 0.237, 0.327} 
                     # the same replicas in the second bin give
                     # ALU = {0.137, 0.149, 0.105} and ALL = {0.183, 0.232, 0.306} 
                - [[0.132, 0.127, 0.140], [0.295, 0.237, 0.327]] 
                - [[0.137, 0.149, 0.105], [0.183, 0.232, 0.306]]
                - [[0.097, 0.113, 0.120], [0.259, 0.262, 0.259]]
            norm_unc: [0.001, 0.002] 
                    # normalisation uncertainties (optional)
                    # ALU +- 0.001, ALL +- 0.002 (for each kinematic bin)
            norm_unc_contrib_label: ["target_pol", "beam_pol"] 
                    # labels of contributions to 
                    # systematic uncertainties (optional)
                    # limited to 40 characters 
            norm_unc_contrib: [[0, 0.001], [0.0014, 0.0019]] 
                    # contributions to systematic
                    # uncertainties (optional)
                    # contributions are: 
                    # 0.0006 and 0.0014 for source labeled as 'target_pol'
                    # 0.0008 and 0.0019 for source labeled as 'beam_pol'

correlation: #definition of correlation matrices used in data file

    - ["corr_matrix1", 1, 0.2, 0.2, 1] 
                    # 2D matrix labeled as 'corr_matrix1', 
                    # with row-by-row values ((1,0.2),(0.2,1))
    - ["corr_matrix2", 1, 0.1, 0.3, 0.2, 0.1, 1, 0.5, 0.1, 0.3, 0.5, 1, -0.2, 0.2, 0.1, -0.2, 1] # 4D matrix
\end{minted}

\begin{figure}[!ht]
\begin{center}
\includegraphics[width=0.2\textwidth]{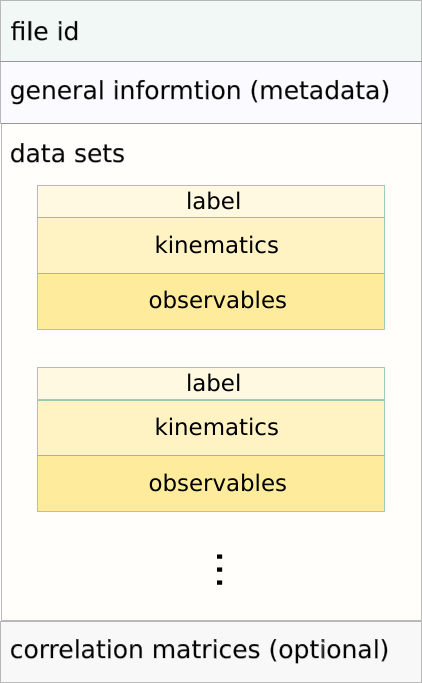}
\end{center}
\caption{General structure of data files.}
\label{fig:structure}
\end{figure}

All data files are made out of three to four sections (\texttt{correlation} section is optional), see Fig.~\ref{fig:structure}. These are:
\begin{itemize}
\item \texttt{uuid}: data file identifier
\item \texttt{general\_info}: general information about data stored (metadata)
\item \texttt{data}: data stored
\item \texttt{correlation}: definition of correlation matrices used throughout \texttt{data} section
\end{itemize}

A single data file can store multiple data sets via \texttt{data\_set} subsections, each one containing:
\begin{itemize}
\item \texttt{label}: data set label
\item \texttt{kinematics}: kinematics stored
\item \texttt{observable}: observables stored
\end{itemize}

The idea behind storing multiple data sets in a single data file is that experiments often use the same data to separately extract observables as functions of independent variables. For instance, in Ref.~\cite{COMPASS:2013fsk}, COMPASS reports the measurement of $A_{\mathrm{UT}}$ asymmetry for deeply virtual meson production in 1D bins of $x_{B}$, $Q^2$ or $p_{T}^2\approx -t$, based on raw data obtained in the same experimental runs. Therefore, in this case, it makes sense to store three different data sets in a single data file. These data sets must be distinguished by unique labels, such as \texttt{"xB\_dep"}, \texttt{"Q2\_dep"} and \texttt{"pT2\_dep"}, so that users may unambiguously choose which data sets (and, therefore, kinematic dependencies) they want to use in their analyses. In the case of lattice-QCD results, many observables are computed on the same set of samples of the QCD vacuum -- the so-called gauge configurations -- and therefore exhibit correlations among themselves. It is useful to group such observables together to allow a proper handling of those uncertainties through replicas.

The presented data file defines just one data set labeled \texttt{"Q2\_dep"}. The stored data are the $A_{\mathrm{LU}}$ and $A_{\mathrm{LL}}$ asymmetries measured in 2D bins of $x_{B}$ and $Q^2$. Three kinematic points are defined, with the mean values of kinematic variables and measured observables, together with uncertainties, specified in Table~\ref{tab:example_points}. In our example, all statistical uncertainties are symmetric. On the other hand, systematic uncertainties are symmetric only for the first point. For the second point, one of the uncertainties is asymmetric. In contrast, for the third point, the uncertainties are correlated according to \texttt{corr\_1} correlation matrix defined in \texttt{correlation} section of the data file. In addition to the statistical and systematic uncertainties, we also have information on normalization uncertainties, which are typically related to the polarization measurements. For a given observable, this type of uncertainty is common to all data points. We note that keeping a detailed track of various types of uncertainties is important for further propagation, as each type of uncertainty is typically handled differently in phenomenological analyses; see, for instance, details of the replication method described in Ref.~\cite{Moutarde:2019tqa}.

\begin{table*}[!ht]
\centering
\caption{Three data points stored in the exemplary data file. In addition to the kinematics (mean values of $x_{B}$ and $Q^2$) and the values of the measured observables, statistical ($\sigma_{\mathrm{stat}}$), systematic ($\sigma_{\mathrm{sys}}$) and normalization ($\sigma_{\mathrm{norm}}$) uncertainties are also provided. The quantities marked with an asterisk are correlated, with a correlation factor of 0.2.}
\begin{tabular}{
C{1.15cm}C{1.15cm}
C{0.3cm}
C{1.15cm}C{1.15cm}C{1.15cm}C{1.15cm}
C{0.3cm}
C{1.15cm}C{1.15cm}C{1.15cm}C{1.15cm}
}
\toprule
\multirow{2}{*}{$x_{B}$} & 
\multirow{2}{*}{$Q^2\,[\mathrm{GeV}^2]$} &&
\multicolumn{4}{c}{$A_{\mathrm{LU}}$} &&
\multicolumn{4}{c}{$A_{\mathrm{LL}}$}\\
& && value & $\sigma_{\mathrm{stat}}$ & $\sigma_{\mathrm{sys}}$ & $\sigma_{\mathrm{norm}}$ && value & $\sigma_{\mathrm{stat}}$ & $\sigma_{\mathrm{sys}}$ & $\sigma_{\mathrm{norm}}$ \\
\midrule
$0.2$ & $1.5$ && $0.13$ & $\pm 0.03$ & $\pm 0.02$\phantom{*}  & $\pm 0.001$ && $0.29$ & $\pm 0.08$ & $\pm 0.01\phantom{*}$ & $\pm 0.002$\\
$0.3$ & $2.2$ && $0.14$ & $\pm 0.04$ & $\pm 0.01$\phantom{*}  & \texttt{"}      && $0.24$ & $\pm 0.07$ & $_{-0.02}^{+0.03}$ & \texttt{"} \\
$0.4$ & $3.1$ && $0.11$ & $\pm 0.03$ & $\pm 0.01$*            &\texttt{"}      && $0.26$ & $\pm 0.02$ & $\pm 0.02$* & \texttt{"}\\
\bottomrule
\end{tabular}
\label{tab:example_points}
\end{table*}

In addition to the basics described so far, we have additional information stored in our exemplary data file. For kinematics, the definition of bins used to extract observables is provided. For instance, observables measured at the first kinematic point were extracted in the following 2D bin: $0.11 < x_B < 0.25$ and $1~\mathrm{GeV}^2 < Q^2 < 2~\mathrm{GeV}^2$. Such information is useful not only for bookkeeping but also allows for more precise comparisons between experimental data and theory. Moreover, in some cases, like in Ref.~\cite{COMPASS:2018pup}, experiments only provide the definition of bins and the cross-section integrated within these bins.

In e.g. lattice-QCD computations, the mean values of kinematic variables and related uncertainties are usually estimated from a set of replicas. Values given by these replicas can be stored in the data files via \texttt{replica} key. In our example, the first replica in the first bin gives $\{x_{B} = 0.2005, Q^{2} = 1.514~\mathrm{GeV}^2\}$, the second $\{x_{B} = 0.1967, Q^{2} = 1.502~\mathrm{GeV}^2$\}, etc. For observables, in addition to the information provided by individual replicas, we may store information on various contributions to the uncertainties. These contributions are distinguished by labels. For instance, as shown in Table~\ref{tab:example_points}, the normalization uncertainty associated with all $A_{\mathrm{LL}}$ points is `0.002'. With additional information provided by \texttt{norm\_unc\_contrib} key, we know that the contribution of beam polarization uncertainty to this value is \texttt{0.0014}, while the contribution of target polarisation uncertainty is \texttt{0.0019}. For $A_{\mathrm{LU}}$, only the beam polarisation uncertainty contributes. 

The values of keys such as \texttt{data\_type}, \texttt{unit}, and \texttt{hadron\_beam\_type} are predefined, resembling the concept of enumerators known in C++. This approach helps maintain database coherence and prevents pollution from inconsistent nomenclature. For instance, the library will process the value \texttt{unit:~xB} without interruption but will trigger an error for \texttt{unit:~xBj}. This feature is also crucial for users, as they can build their code around the library without needing to create a separate dictionary for each data file. The enumerators are defined in separate YAML files, making it easy to add new values. For example, predefined data types, which users must specify using the \texttt{data\_type} key, are defined in the following file:
\begin{minted}[
    fontsize=\footnotesize,
    % linenos,
    % frame=leftline,
    % framesep=5mm
  ]{yaml}
---
data:

  # structure function (like elastic FFs, F2, etc.)
  - name: STRUCTURE_FUNCTION
    description: "structure function"
    required_name: [ hadron_type ]
    required_type: [ particle ]
  # elastic
  - name: LATTICE_QCD
    description: "lattice QCD"
    required_name: [ hadron_type, pion_mass, lattice_spacing ]
    required_type: [ particle, float, float ]
  # exclusive: DVCS
  - name: DVCS
    description: "deeply virtual Compton scattering"
    required_name: [ lepton_beam_type, lepton_beam_energy, hadron_beam_type, hadron_beam_energy ]
    required_type: [ particle, float, particle, float ]
\end{minted}
This example also demonstrates another feature of the database library, namely, type checks. For instance, \texttt{data\_type:DVCS} requires defining \texttt{lepton\_beam\_type}, \texttt{lepton\_beam\_energy} and \texttt{hadron\_beam\_type} in the \texttt{conditions} section. However, not just any data can be specified there. For example, \texttt{lepton\_beam\_energy} must be a float value, while \texttt{lepton\_beam\_type} must be a particle type, such as \texttt{p} or \texttt{H4}, that is recognized by the ``particle'' library~\cite{particle}. Providing an incorrect data type will result in an error triggered during the processing of such a corrupted data file. Several additional checks are implemented in the library. For instance, when defining an observable, one must specify the associated unit type. The predefined unit types include, for instance, \texttt{EVm2}, which applies to all units that can be converted to $\mathrm{eV}^{-2}$ in the natural system, such as barns and $\mathrm{GeV}^{-2}$, and \texttt{ANGLE}, applicable to radians, degrees, etc. This solution prevents setting, for example, $Q^{2} = 2~\mathrm{GeV}$, and also opens the possibility of implementing a unit conversion at a later stage of the project. It is worth noting that the use of the ``particle'' library allows easy access to information such as mass and charges of particles, which can be straightforwardly used in users' analysis codes. 

Alphanumeric data, such as those associated with the \texttt{collaboration} or \texttt{label} keys, are limited to a certain number of characters (see comments in the example data file). This solution addresses practical concerns of storage, performance and data integrity, and is routinely implemented in SQL-like databases. Since we foresee using an SQL format in the future as an alternative to YAML files stored on GitHub, it seems appropriate to comply with SQL practices now. Aside from the length check, alphanumeric data are not validated in any other way. It is, therefore, up to the users and admins to insert meaningful data of this type. Additional information to help understand the content of the data file can be inserted as comments, i.e., via the \texttt{comment} key or the '\#' YAML tag. This may be particularly helpful in deciphering labels set via the \texttt{label}, \texttt{sys\_unc\_contrib\_label}, and \texttt{norm\_unc\_contrib\_label} keys. For the \texttt{uuid} key, serving as an identifier of the data file, we intend to use short universally unique identifiers, which can be easily generated in Python or even dedicated web-pages, like~\cite{uuidgenerator}. 

We stress that, if necessary, the data format can be modified in the future to include additional information while maintaining full backward compatibility. In particular, for lattice QCD, we highlight the possibility of adding indicators of data quality. Criteria for this purpose have been developed over many years by the Flavour Lattice Averaging Group (FLAG)~\cite{FlavourLatticeAveragingGroupFLAG:2024oxs}, and could be adopted for the case of structure functions, taking into account specific challenges unique to pseudo- and quasi-distribution approaches, such as solving the inverse moment problem. The introduction of such criteria and a transparent method for assigning them would facilitate comparisons between calculations performed under various conditions and approaches, enabling a meaningful global analysis of lattice-QCD results.


\section{Users interface}
\label{sec:ui}

A library is provided to check the database content, load a given data file, and access stored information. The library is written in Python, allowing for its straightforward use in analysis codes. A C++ wrapper for this library, based on the generic Python.h library, is also available. This enables the database to be used in projects like GeParD~\cite{gepard} and PARTONS~\cite{Berthou:2015oaw}, which are written in Python and C++, respectively. The Python library is available in the PyPI repository, making its installation particularly easy -- typically, the installation is reduced to executing only one command:
\begin{minted}[
    fontsize=\footnotesize,
    % linenos,
    % frame=leftline,
    % framesep=5mm
  ]{bash}
pip3 install gpddatabase
\end{minted}
The code of C++ wrapper is available in the main repository of the project~\cite{database_repo}, and it requires the Python library to be pre-installed. The wrapper can be easily incorporated in any C++ projects thanks to the provided CMake~\cite{cmake} module.

The following example demonstrates how to access the database using the Python library. For simplicity, we present only the most basic operations for the Python interface. Examples, including those for C++, and technical documentation are available on the project's main page.
\begin{minted}[
    fontsize=\footnotesize,
    % linenos,
    % frame=leftline,
    % framesep=5mm
  ]{python}
# import module
import gpddatabase

# make a reference to the database
db = gpddatabase.ExclusiveDatabase()

# print availible uuids
print(db.get_uuids())

# load a given data file
ob = db.get_data_object('9j7gof4d')

# print date of insertion stored in general_info section
print(ob.get_general_info().get_date())

# print labels of availible datasets
print(ob.get_data().get_data_set_labels())

# print number of points stored in 'Q2_dep' dataset
print(ob.get_data().get_data_set('Q2_dep').
   get_number_of_data_points())

# make a referenece to the first point
point = ob.get_data().get_data_set('Q2_dep').
   get_data_point(0)

# print names of kinematic variables, then units and values
print(point.get_kinematics_names())
print(point.get_kinematics_units())
print(point.get_kinematics_values())

# print names of observables, then units and values
print(point.get_observables_names())
print(point.get_observables_units())
print(point.get_observables_values())
\end{minted}

A local installation of the library involves downloading all data files available in the database. The location of these files can be checked in the following way.
\begin{minted}[
    fontsize=\footnotesize,
    % linenos,
    % frame=leftline,
    % framesep=5mm
  ]{python}
# import module
import gpddatabase

# make a reference to the database
db = gpddatabase.ExclusiveDatabase()

# print location of data files
print(db.get_path_to_databse())
\end{minted}
This is a straightforward solution that simplifies the use of the database (no need to manually select and download the files) and allows, for instance, easy addition and testing of new content. As already mentioned in Sect.~\ref{sec:basics}, the current size of the library is only a few megabytes but will grow in the future. Therefore, switching to more elaborate ways of managing data files, such as those implemented in the LHAPDF library~\cite{Buckley:2014ana}, is an option to be considered for future releases of the project.


\section{Summary}
\label{sec:summary}

This article summarizes the key concepts behind the open database we propose for use in the GPD community. This lightweight database is well suited for GPD phenomenology and is designed to store both experimental and lattice-QCD data. It can also be used to benchmark GPD-related developments, such as GPD models. The database utilizes a new data format based on the YAML serialization language, which enables storage of essential information for modern phenomenology analyses, including replica values, bin sizes, correlation matrices, and more. Predefined types are used to ensure data consistency. A user interface is provided, allowing straightforward integration with analysis codes in Python and C++. Additional technical details on the project are available on the project's main page~\cite{database}.

\begin{acknowledgements}
The authors are grateful to K. Kumerički, S. Liuti, W. Melnitchouk, E. Moffat and A. Prokudin for valuable discussions. This material is based in part upon work supported by the U.S. Department of Energy, Office of Science, Office of Nuclear Physics under contract DE-AC05-06OR23177. The research described in this paper was conducted in part under the Laboratory-Directed Research and Development Program at Thomas Jefferson National Accelerator Facility for the U.S. Department of Energy. This work has been supported in part, by l’Agence Nationale de la Recherche (ANR), project ANR-23-CE31-0019. For the purpose of open access, the author has applied a CC-BY public copyright licence to any Author Accepted Manuscript (AAM) version arising from this submission. K.~C.\ is supported by the National Science Centre (Poland) grant OPUS No.\ 2021/43/B/ST2/00497. M.~C. acknowledges financial support by the U.S. Department of Energy, Office of Nuclear Physics under Grant No.\ DE-SC0025218. The work of P.~S.
was supported by the Grant No.~2024/53/B/ST2/00968
of the National Science Centre, Poland.
\end{acknowledgements}

\bibliography{main}
\bibliographystyle{spphys}

\end{document}